\documentclass[aps,prl,floatfix,twocolumn,superscriptaddress]{revtex4-1}
\usepackage{latexsym}
\usepackage{graphicx}
\usepackage{rotating}
\usepackage[colorlinks=true,linkcolor=blue,citecolor=red]{hyperref}
\usepackage{amsmath,amssymb,amsfonts}
\usepackage{bm}
\usepackage{color}

\newcommand{\emphasize}{\emph}

\def\onlinecite#1{\cite{#1}}

\newcommand{\norm}[1]{\ensuremath{| #1 |}}

\usepackage{natbib}
\usepackage[compact]{titlesec}
\titlespacing{\section}{0pt}{*0}{*0}
\titlespacing{\subsection}{0pt}{*0}{*0}
\titlespacing{\subsubsection}{0pt}{*0}{*0}
\begin{document}

\title{Augmented hybrid exact-diagonalization solver for dynamical mean field theory}

\author{C.~Weber$^{*}$}
\affiliation{Cavendish Laboratory, J.J. Thomson Ave, Cambridge CB3 0HE, U.K.}
\author{A.~Amaricci}
\affiliation{CNR-IOM, SISSA, Via Bonomea 265, 34136 Trieste, Italy.}
\author{M.~Capone}
\affiliation{CNR-IOM, SISSA, Via Bonomea 265, 34136 Trieste, Italy.}
\affiliation{Physics Department, University "Sapienza", Piazzale A. Moro 2, 00185 Rome, Italy.}
\author{P.~B.~Littlewood}
\affiliation{Cavendish Laboratory, J.J. Thomson Ave, Cambridge CB3 0HE, U.K.}
\affiliation{Physical Sciences and Engineering, Argonne National Laboratory, Argonne, Illinois 60439, U.S.A.}


\begin{abstract}
We present a new methodology to solve the Anderson impurity model, in the context of dynamical mean-field theory, based on the exact diagonalization method.
We propose a strategy to effectively refine the exact diagonalization solver by combining a finite-temperature Lanczos algorithm with an adapted version of the cluster perturbation theory. 
We show that the augmented diagonalization yields an improved accuracy in the description of the spectral function of the single-band Hubbard model
and is a reliable approach for a full d-orbital manifold calculation. 
\end{abstract}

\maketitle

The understanding of materials with strongly correlated electrons is one of the main challenges of modern solid state physics. 
Triggered by the discovery of high-temperature superconductivity in copper-oxides, 
the study of doped Mott insulators has grown in the last decades, building on the development of theoretical tools designed 
to solve accurately models of strongly correlated electrons.  

Despite that the exact solution of simple correlated theoretical models in two or three dimensions is lacking, 
accurate predictions for the properties of strongly correlated solids are obtained by using approximations \cite{kotliarRMP}. 
A central role has been played by the dynamical mean field theory (DMFT)\cite{rmp}, a nonperturbative method which allowed for the first complete 
description of the Mott-Hubbard transition. This method has been extended to a variety of correlated methods and combined with density functional theory, 
leading to remarkable agreement with the properties of many correlated materials. 

Within DMFT the original lattice problem is mapped onto an effective Anderson Impurity Model (AIM), 
which can be solved numerically. The numerical solution of the AIM is the real bottleneck of a DMFT calculations, 
and the development of \emph{impurity solvers} which are both accurate and computational-wise efficient is a very active line of research.
Among the many solvers proposed and extensively tested through the last decade, those based on exact diagonalization (ED) \cite{caffarel_edstar,lucaED,ed_solver_liebsch_review} or on the continuous-time quantum Monte Carlo (CTQMC) \cite{wernerctqmc,*Haule07,*Gull2011} are used extensively in this line of research.

Indeed, the recent development of CTQMC has generated a strong activity in the field. For single-site DMFT, CTQMC yields an exact solution of the AIM problem 
within the statistical error bars in imaginary time.  The main limitation of the approach is that the evaluation of real-frequency spectra requires a poorly conditioned analytical continuation, 
based on the maximum entropy method\cite{mem,*Gunnarsson10} (or some alternative strategy), which strongly limits the possibility to study fine details of the spectra. 
For multi-orbital or cluster extensions of DMFT, CTQMC suffers from the \emph{fermionic sign-problem} 
as long as inter-orbital hybridizations are present, and it is therefore limited to finite temperatures.

The ED solvers are instead based on a finite discretization of the AIM, through the representation of the effective bath in terms of a small number of \emphasize{bath-sites}.
In practical implementations the bath size ($N_b$) is severely limited because of the growth of the Hilbert space: its dimension scales exponentially with
the total number of sites $N_s$ (bath sites and impurity orbitals). Nonetheless, the use of Lanczos-based algorithms 
allows to deal with large Hilbert spaces, and the discretization at low temperature\cite{lucaED,leibsch_ed_lanczos_dmft} 
is fine enough to accurately compute the thermodynamic and static observables.
In particular, the finite size effects affect the spectral functions, which are slowly converging to the continuous features of the exact DMFT solution.
Notwithstanding reasonably accurate static phase diagrams obtained with the Lanczos solver for three\cite{Capone02,ed_solver_liebsch_review} and five\cite{Ishida10} orbitals system, 
the limitation in the bath size becomes also particularly relevant for multi-orbital AIM models,
which are necessary to realistically describe transition metal oxides, as the impurity's orbitals (five for a d manyfold) contribute to the enlargement of the Hilbert space.

In the view of the increasing interest in the effect of multi-orbital and multi-site Coulomb correlations in transition metal oxides, 
an improvement of the ED methodology is hence called for. 

In this work we propose a simple, yet efficient, modification of the ED solver for the DMFT, which allows to treat larger bath sizes without increasing the computational cost. In particular, we propose a combination of ED, which is used to solve a small part of the bath, but we include an extra set of bath levels which are perturbatively 
coupled to the first part using concepts along the line of cluster perturbation theory (CPT) \cite{tremblay_cpt_square_lattice_pseudogap,cpt_senechal,*Senechal02}. 
The augmented effective bath in the ED-CPT method improves the accuracy in both static and spectral properties as compared with standard ED, 
and it allows for accurate calculations for multi-orbital models. 

\begin{figure}
\begin{center}
\includegraphics[width=0.8\columnwidth]{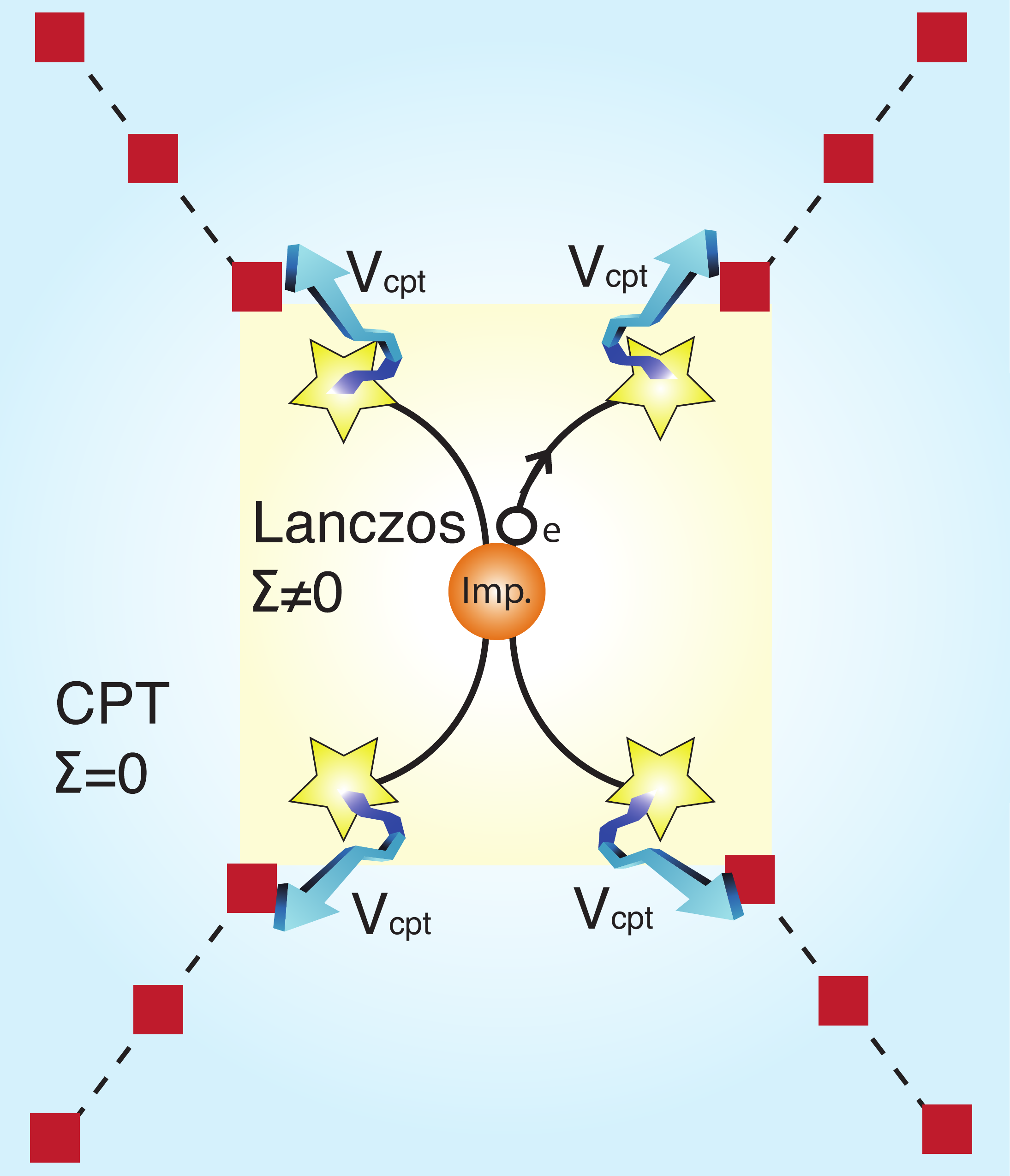}
\caption{
(Color online) Sketch of the AIM in ED-CPT: the impurity (center circle) is connected to $N_b$ nearest neighbor bath sites (stars),
and each of the latter is connected by a small perturbative parameter $V_{CPT}$ to a line of additional sites (squares, $N_{c
}$ sites in each chain). The inner system (light filling) is solved by using a standard Lanczos algorithm. The Green's function of the full system (inner and outer region) is obtained by treating the full system by cluster perturbation theory.}
\label{fig1}
\end{center}
\end{figure}

Within DMFT, the effective AIM is subject to a self-consistency condition which relates the Green's function of the impurity model $G(i\omega_n)$ to the so-called Weiss field ${\cal{G}}_0^{-1}(i\omega_n)$, which completely characterizes the AIM. For the single band case, the Hamiltonian of the AIM reads:
\begin{multline}
\label{aim}
H=\sum\limits_{p\sigma}{\epsilon_{p\sigma}a^\dagger_{p\sigma}a_{p\sigma}} 
+\sum\limits_{p\sigma}{V_{p\sigma}(a^\dagger_{p\sigma}d_{\sigma}+hc)} + U \hat n_{\uparrow} \hat n_{\downarrow}
\end{multline}
where $a^\dagger_{p\sigma}$ ($a_{p\sigma}$) creates (destroys) a particle with spin $\sigma$ in the p-orbitals ($p \in [1,N_p]$) of the bath and $d^\dagger_{\sigma}$ ($d_{\sigma}$) creates (destroys) a spin $\sigma$ particle on the impurity, and $U$ is the static Coulomb repulsion on the impurity. Spin indices are omitted throughout this letter.  
For a fixed set of the Hamiltonian parameters, we solve the AIM (\ref{aim}) by using a Lanczos algorithm to converge the low-lying states of the spectrum\cite{lucaED,ed_solver_liebsch_review} which contribute to the thermal average:  we consider an energy cutoff $E_\mathrm{max}$ such that the Boltzman weight of the discarded states is negligible $e^{-\beta(E_\mathrm{max}-E_0)}<0.001$, where $E_0$ is the ground state energy. Once the eigenstates are obtained we compute the dynamical and static observables\footnote{A low temperature $T/D=0.01$ is used in all the present calculations}. 

In the DMFT methodology, the Weiss field ${\cal{G}}_0^{-1}(i\omega_n) = i\omega_n +\mu - \Delta(i\omega_n)$, obtained through the mean-field self-consistency condition, must be represented in the discrete basis of the AIM. To this intent, we minimize a suitable distance between the Weiss field (or, equivalently, the function $\Delta$) and the discretized AIM hybridization function $\Delta^\textit{\tiny ED}(i\omega_n)=\sum_{p=1}^{N_b} {V_{p}^2}/({i\omega_n - \epsilon_{p}})$:
\begin{equation}
\label{distance}
d = \sum\limits_{\omega<\omega_0} {\left|{ \Delta^\textit{\tiny ED} \left({i{\omega }_{n}}\right)  -{\Delta \left({i{\omega }_{n}}\right)} }\right|^2}/{\omega_n},  
\end{equation}
where $\omega_0$ is a hard cutoff on the number of Matsubara frequencies (we set $\omega_0=100$ throughout the rest of this work). The denominator is a weighting factor that can be used to improve the precision on the low energy frequencies close to the metal-insulator transition \cite{Capone04,ed_solver_liebsch_review}. The distance $d$ is furthermore renormalized by the number of frequencies $\omega<\omega_0$. 

The minimization (\ref{distance}) boils down to the problem of fitting an arbitrarily complicated function with a finite number of rational functions. This is the only source of potential systematic errors and limitations of the standard ED algorithm. Indeed, for low temperatures the number of bath-orbitals required to resolve the low energy features characterizing the hybridization function can be large.
Noteworthy, there is no simple limit where the fitting method used in the ED solver is exact, in particular the finite size effects and systematic errors in the fitting procedure are equally severe for the
trivial $U=0$ and large $U$ limits.  
Moreover, the fitting procedure may become problematic near the zero repulsion $U \approx 0$, 
where the hybridization might have a complicated structure difficult to capture with the fitting method. 
It is worth reminding that, while there is no strong limitation to the number of bath-orbitals at the level of the fit, the size of the Hilbert space of the effective AIM is exponentially growing with the number of sites, rapidly becoming out of reach for any ED-based algorithm. 
Indeed, although the Lanczos solver yields static and dynamical observables with a remarkable precision, the 
effective discretized AIM might very poorly reproduce the DMFT hybridization, giving in turn large errors in the spectral functions.

In this work, we solve this problem by extending the AIM to a larger system. This extension is realized by coupling additional lines of $N_c$ non-interacting sites to each of the $N_b$ bath-orbitals (see Fig.~\ref{fig1}). 
In the spirit of the CPT, the couplings $V_{p,\textit{\tiny CPT}}$ between each chain and the corresponding bath-site are introduced as small perturbative parameters\cite{cpt_senechal,*Senechal02}. Indeed, we treat the hybridization between the AIM and the chains within CPT, which amounts to use the following expression for the impurity Green's function:
\begin{equation}
\bold{G}^{CPT}(i\omega_n)=\frac{1}{\bold{G}^{V_0}(i\omega_n)^{-1}-\bold{V_{CPT}}},
\end{equation}
where $\bold{G}^{CPT}(i\omega_n)$ is the full Green's function of the coupled (AIM+chains) system, $\bold{G}^{V_0}(i\omega_n)$ is a block-diagonal matrix which contains the Green's function of the decoupled AIM and chains, and $\bold{V_\text{\tiny CPT}}$ is the hybridization matrix that connects the two subsystems. All the matrices are defined in the space composed by the bath-orbitals and the chains. The hybridization function of the ED-CPT explicitly depends on the additional CPT parameters, and can be written in orbital space as $\mathbf\Delta^\textit{\tiny ED} \left({i{\omega }_{n}}\right) =\mathbf{V}^\dagger \left(  i\omega_n -  \mathbf{\epsilon} \right)^{-1} \mathbf{V}$ \cite{senechal_cpt_note_explanations}, where $\mathbf{V}$ and $\mathbf{\epsilon}$ are respectively the matrices of the impurity-bath and bath-bath couplings. 
Thus, using this simple strategy, we are able to effectively increase the size of the bath and to improve our fitting procedure without increasing the size of the Hilbert space in the Lanczos diagonalization. 
The CPT is exact both in the $U=0$ limit and in the infinite-$U$ limit (in which all the hybridizations are negligible)\cite{cpt_senechal,*Senechal02}, while for intermediate value of the interaction, the method is expected to be reliable provided that $\bold{V_{CPT}}$ remains small. Thus, we add a simple Lagrange term $\lambda \left(  \sum_p{\norm{V_{p, \text{\tiny CPT}}}^2 }\right)$ to the fitting distance $d$ and impose an upper hard-cutoff on the coupling $V_{p,\text{\tiny CPT}}$ \footnote{We used $\lambda=10^{-6}$ and a cutoff of $0.1$D.}.  We emphasize that the present method is easily implemented, since it only differs from the standard ED algorithm in the fitting procedure.

\begin{figure}
\begin{center}
\includegraphics[width=1.0\columnwidth]{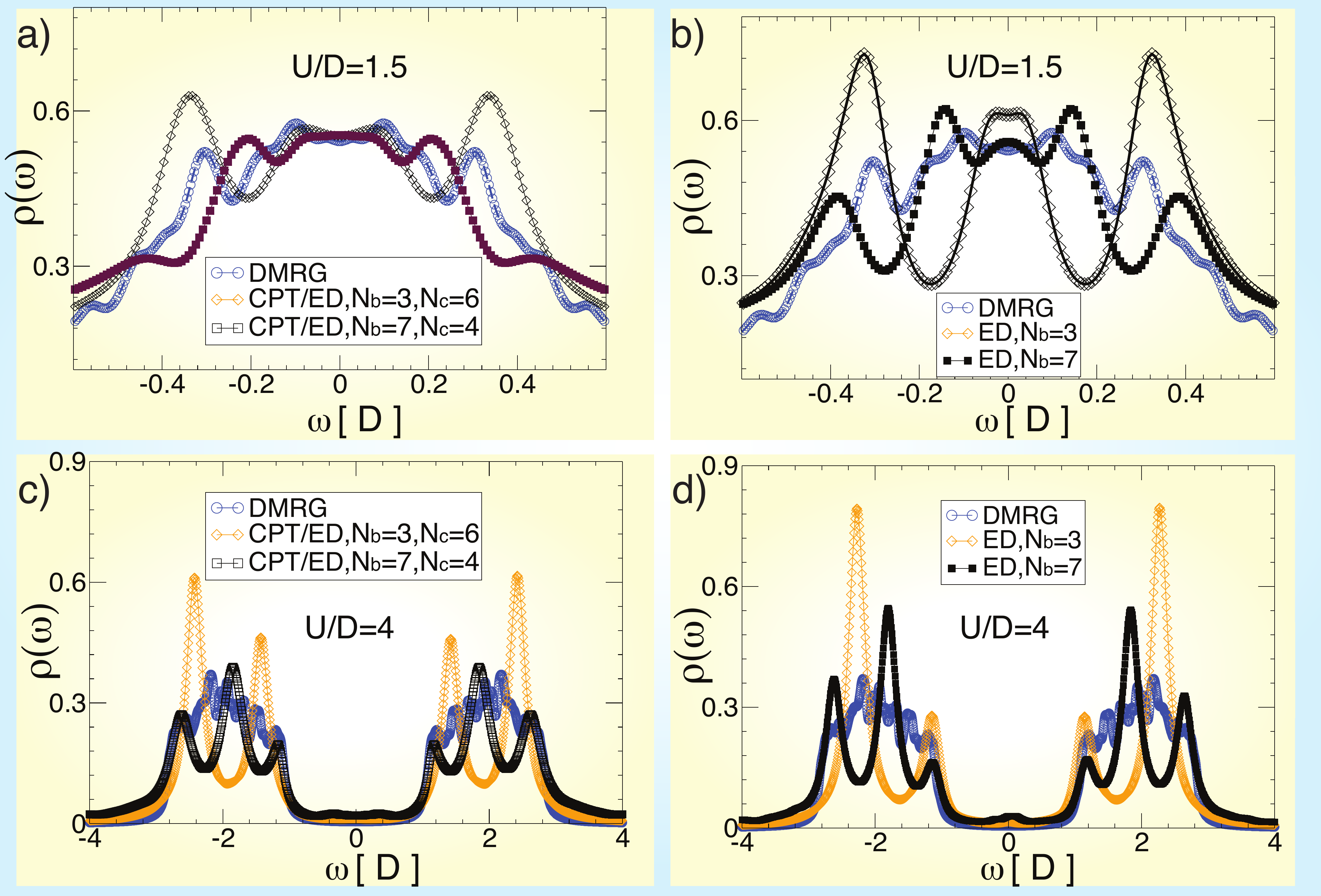}
\caption{
(Color online) Spectral function $\rho(\omega)$ obtained for the Hubbard model solved by ED-CPT at a) $U/D=1.5$ and c) $U/D=4$. For comparison, the spectral functions obtained by ED at b) $U/D=1.5$ and d) $U/D=4$ are shown. The ED results with $N_b=3$ (diamonds) and $N_b=7$ (squares) bath sites are compared with the DMRG results (circles). The CPT-ED results obtained with $N_b=3$ inner bath sites and augmented with lines of $N_{c}=6$ sites (diamonds) and respectively $N_b=7$ and $N_{c}=4$ (squares) compare remarkably well to the DMRG data (circles).}
\label{fig2}
\end{center}
\end{figure}

To begin with, we benchmark the ED-CPT method for the single-band half-filled Hubbard model against the standard ED algorithm and a density matrix renormalization group (DMRG) calculations \cite{white_dmrg,*dmrgRMP,*daniel1,*Sordi2009}. 
For sake of simplicity, we work on the Bethe lattice with a semi-elliptical  density of states $\rho(\epsilon)=\pi D^2 \sqrt{D^2-\epsilon^2} / 2$. For this system the self-consistency condition simplifies  
to a linear relation $\Delta(i\omega_n)= G(i \omega_n) \left( D^2 / 4 \right) $~\cite{OLD_GABIS_REVIEW}. 

We first compare the spectral function $\rho(\omega) = -\text{Im}G(\omega)/\pi$ obtained by ED-CPT (see Fig.~\ref{fig2}.\textbf{a,c}) with the one obtained using ED with the same number of bath orbitals (see Fig.~\ref{fig2}.\textbf{b,d}). The results are benchmarked against DMRG calculations performed using a large number of bath sites $N_b=30$ ($U=1.5$) or $N_b=60$  ($U=4$). 
As expected, for a small number of bath sites $N_b=3$ we observe strong deviations for $\norm{\omega}/D<0.2$ in the ED results at small repulsion $U/D=1.5$ (Fig.~\ref{fig2}.\textbf{b},diamonds). For this interaction, the system is in a metallic state, and hence a dense discretization of the hybridization is required to recover the right low-energy spectral properties. This is dramatically improved in ED-CPT (Fig.~\ref{fig2}.\textbf{a},diamonds), where the spectral function with $N_b=3$ is essentially identical to the DMRG results in the low-energy range. Larger-energy features, such as the peaks at $\norm{\omega}/D= \pm 0.35$ are also improved by ED-CPT. 
For a larger number of bath-orbitals, $N_b=7$, both ED and ED-CPT give results close to the DMRG. Noteworthy, the Lagrange parameter $\lambda$ suppresses the coupling with the chain $\bold{V_{CPT}}$ when the number of bath orbitals is already sufficient to obtain a faithful representation of the hybridization function.
Although the ED-CPT method is expected to provide a significant improvement for small $U$, we now explore the strong interaction regime $U/D=4$ which corresponds to a Mott insulating solution (see Fig.~\ref{fig2}.\textbf{c,d}). The finite discretization of the hybridization function leads to large peaks in the DOS, as seen in the ED results for $N_b=3,7$  (Fig.~\ref{fig2}.\textbf{d}).
ED-CPT indeed smears significantly those artificial structures (Fig.~\ref{fig2}.\textbf{c}) yielding a solution closer to the DMRG result. We note that also the spectral weight at the gap edge is quite improved by ED-CPT. 

\begin{figure}
\begin{center}
\includegraphics[width=1.0\columnwidth]{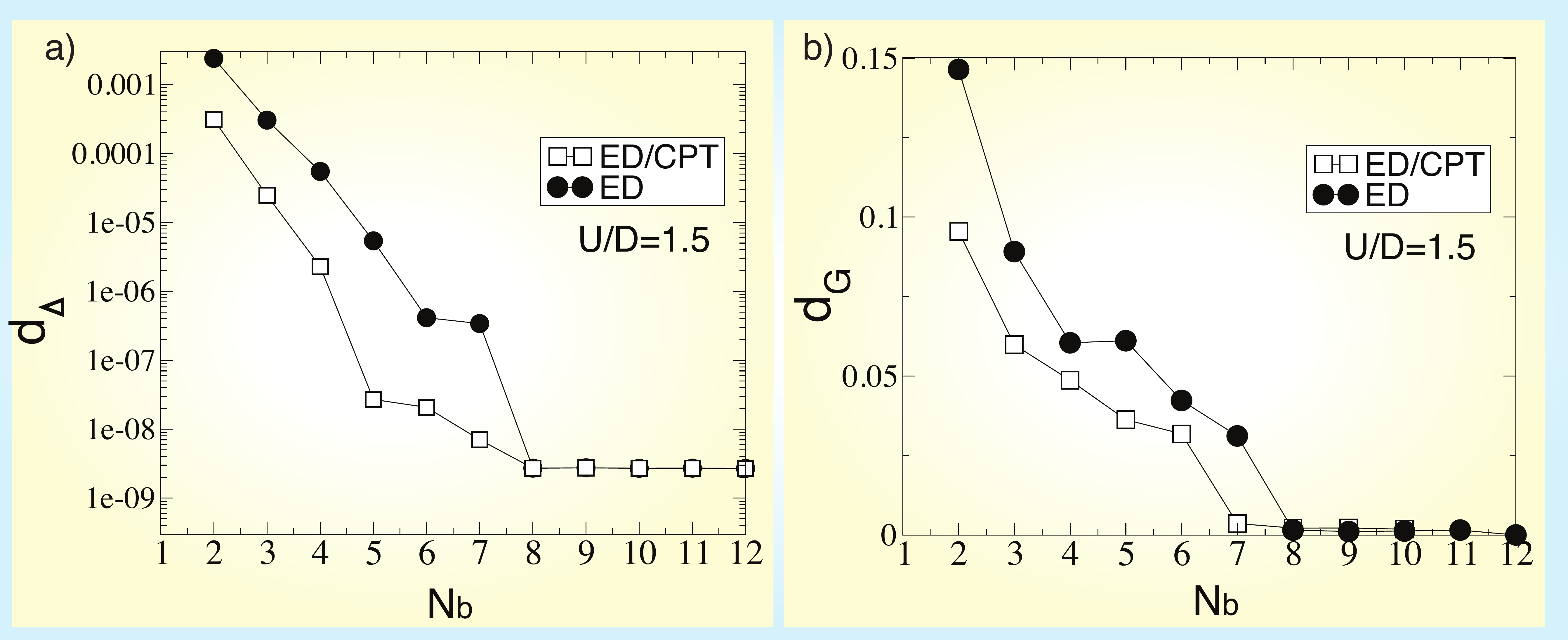}
\caption{
(Color online) a) Fitting distance $d$ obtained by using the ED and ED-CPT. b) Integrated absolute value of the difference  in the spectral functions for the Hubbard model with $U/D=1.5$.}
\label{fig3}
\end{center}
\end{figure}

In order to better assess the improvement of the ED-CPT method over the ED we show  in Fig.~\ref{fig3}.\textbf{a} the fitting distance $d$ as a function of the effective bath size. In particular, we find that at weak interaction $U/D=1.5$ the distance $d$ is an order of magnitude smaller for the ED-CPT than the ED method, up to discretizations $N_b<8$. 
This is especially interesting, since the upper limit $N_b=7$ corresponds to largest system which can be carried out with the full ED. Thus, the ED-CPT can significantly improve also the full-diagonalization DMFT solver, allowing to perform reliable calculations at every temperature.

In Fig.~\ref{fig3} we also report the frequency integral of the modulus of the difference between the spectral function at various values of $N_b$ and the result for the large value $N_b=12$, $d_G=\int_{-\infty}^{\infty}{\left| \rho(m,\omega)-\rho(m=12,\omega) \right| d\omega}$
The behavior of this quantity confirms the significant improvement of the ED-CPT solution with respect to the ED results, especially for $N_b<4$. 
Also for this quantity, for larger number of bath sites ($N_b>8$), no significant difference between ED and ED-CPT is found for $U/D=1.5$. 
The benchmark of the ED-CPT against ED and DMRG results for the Hubbard model demonstrates that when only few bath-orbitals are available to the AIM, the ED-CPT provides a very robust way to improve the ED-DMFT solution, yielding a systematic reduction of the finite size effects at a very moderate computational cost. 

\begin{figure}
\begin{center}
\includegraphics[width=1.0\columnwidth]{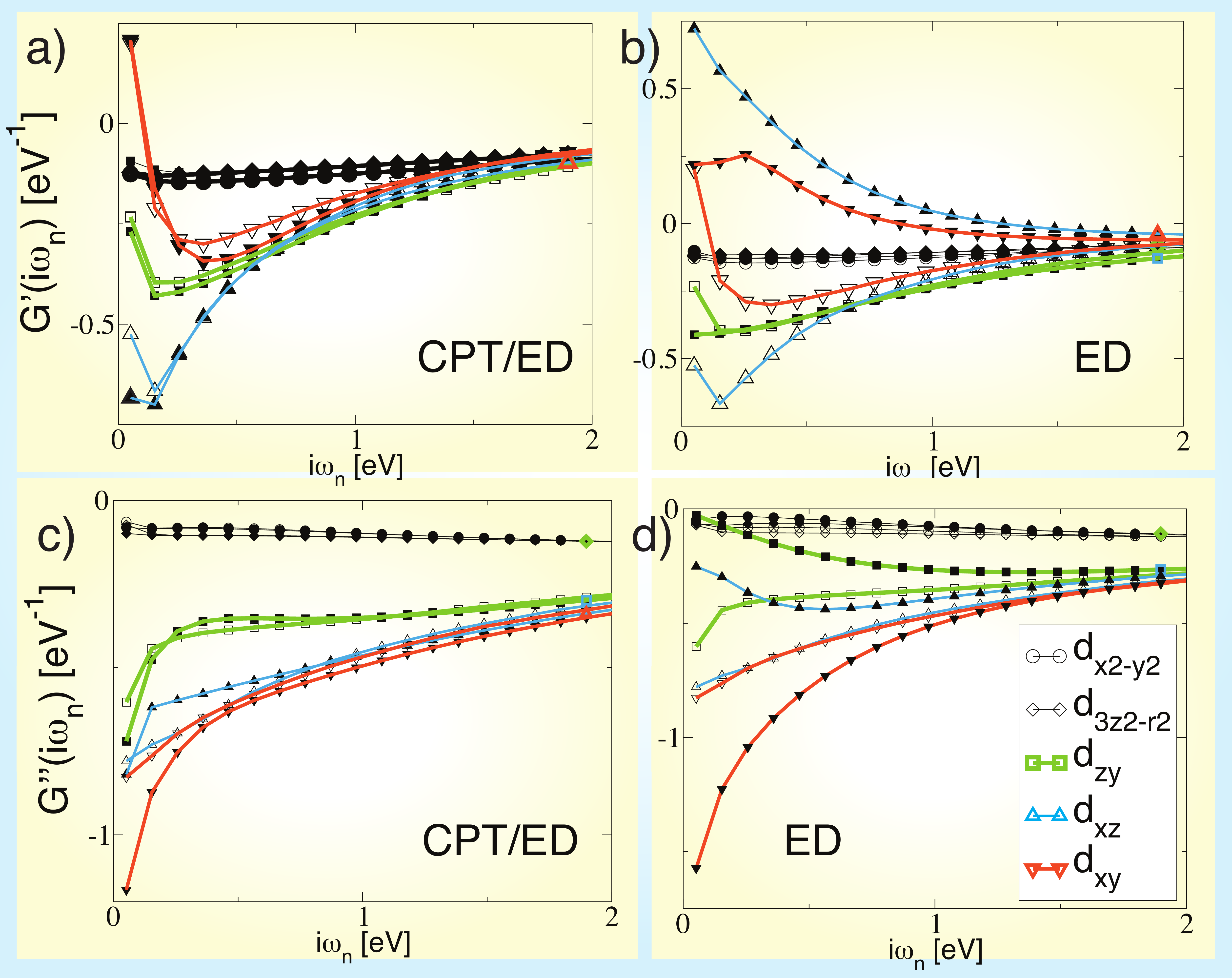}
\caption{
(Color online) Imaginary-frequency Green's function:  Real (a,b) and imaginary part (c,d) of the Green's function $\bold{G}$ for a multi orbital system (paramagnetic VO$_2$).  The data obtained by (a,c) ED-CPT (filled symbols) and (b,d) by ED (filled symbols) are compared with reference data obtained by CTQMC (open symbols) reproduced from Ref.~\cite{our_paper_vo2}. The calculations include all five d orbitals ($e_g$ and $t_{2g}$).}
\label{fig4}
\end{center}
\end{figure}

Having benchmarked the validity of the ED-CPT algorithm against other methods in simple problems, we now turn our attention to multi-orbital systems.
When applied to realistic materials calculation the finite-size effects in the ED-DMFT solution become particularly severe. In particular only few bath sites become available for each impurity orbital, introducing a severe limitation of the method already for the case of d-orbitals.
In order to test the improvements brought in by the ED-CPT method we apply this method to the Vanadium Dioxide \onlinecite{vo2_paper_ref2}.
VO$_2$ is a transition-metal oxide with an open $d$-shell, and hence it requires an impurity model with $5$ impurity levels\cite{dmft_compounds_vo2,*tomczak}. This system is moderately correlated metal  ($U_d=4$eV) with a significant Hund's coupling $J=0.68$eV.  We consider a bath discretization for this system with $N_b=6$ ($N_s=11$). 

In figure Fig.~\ref{fig4} we compare the impurity Matsubara Green's function obtained by the ED-CPT (Fig.~\ref{fig4}.\textbf{a,c}, filled symbols) and by the ED (Fig.~\ref{fig4}.\textbf{b,d}, filled symbols) against the CTQMC results, reproduced from Ref.~\onlinecite{our_paper_vo2} (open symbols). 
We find with ED that the density at the Fermi level 
$\rho(\epsilon_F)=-\text{Im}G(\omega=0)/\pi$ of the d$_{xz}$ and d$_{yz}$ orbitals is quenched, and the d$_{xy}$ has
a very large weight at the Fermi level (Fig.~\ref{fig4}.\textbf{d}). This differs from the CTQMC data, which show that all the t$_{2g}$ orbitals
are contributing to the spectral function at the Fermi level with a weight of the same order. The ED solver is hence unable to capture
the relevant ingredients for this 5 orbital system. We also  tried different normalization in equation (\ref{distance}) as well as more general fitting functions\cite{Koch08}, but we could not significantly improve the results. 

However, by using the extended ED-CPT Lanczos solver, we could recover the main
features at the Fermi level (Fig.~\ref{fig4}.\textbf{c}). In particular, despite that the ED-CPT slightly over-estimate the density of the d$_{xy}$
orbital, the order of the respective contribution of each orbitals is conserved. Another failure of the ED solver is the low energy
frequency behavior of the real part of the Green's function (Fig.~\ref{fig4}.\textbf{b}). 
Indeed, the real part of the Green's function vanishes for particle-hole symmetric situation,
while its positive when the spectral weight below $\epsilon_F$ has largest weight and vice versa.
The CTQMC data show that the most occupied orbital is the d$_{xy}$, and the ED data instead occupies strongly
the d$_{xz}$ orbital instead. This problem is again cured in the ED-CPT (Fig.~\ref{fig4}.\textbf{a}), which recovers 
remarkably the same features as in the CTQMC. The advantages of ED-CPT over CTQMC is that the real frequency observables
are readily available from the calculations without further approximations such as the analytical continuation used in CTQMC.

In conclusion, we have presented a new solver for dynamical mean-field theory, which systematically improves the resolution of low energy scales.
This so-called \emphasize{augmented hybrid exact-diagonalization} solver is based on the exact-diagonalization scheme, which relies
on a finite discretization of the effective bath. 
This new solver systematically reduces the finite size effects of the ED method by 
considering a system of an increased size,
and by treating the additional bath-levels within cluster perturbation theory. 
While this method requires slight changes in the implementation of the self-consistency condition, it introduces no further computational cost with respect to the ED solver. 
We showed that the ED-CPT solver provides a significant and systematic improvement over the ED for the one-band Hubbard model. We extended the calculations to a multi-orbital realistic material (VO$_2$), where the number of impurity orbitals needed to describe the $\text{d}^5$ levels prevents using a large number of bath sites in the ED and yields severe finite size effects. 
In this case we showed that the ED-CPT solver is able to qualitatively improve the ED results and to achieve a remarkable agreement with CTQMC data. The methodology described in this work can easily be extended to any ordered phases, such as magnetism or superconductivity\cite{Kancharla08}.
The CTQMC code used in Ref.~\onlinecite{our_paper_vo2} was written by K.~Haule\cite{Haule_long_paper_CTQMC}.
C.W. was supported by the Swiss National Foundation for Science (SNFS).
P.B.L is supported by the US Department of Energy under FWP 70069. M.C. and A.A. acknowledge financial support from the European Research Council under 
FP7 Starting Independent Research Grant n.240524.

\bibliographystyle{apsrev4-1}
\bibliography{bibliografia,biblio_cedric}
\end{document}